\documentclass[12pt]{article}
\usepackage{amsmath,verbatim,amssymb,epsfig,color,lscape,bm}
\usepackage{graphicx}

\usepackage{soul}
\usepackage{natbib} 
\usepackage{url} 

\usepackage{soul}
\newcommand{\blind}{0}

\begin{document}

\def\spacingset#1{\renewcommand{\baselinestretch}%
{#1}\small\normalsize} \spacingset{1}


\if0\blind
{
  \title{\bf Tolerance Intervals Using Dirichlet Processes}
  \author{Seokjun Choi\thanks{
   The authors gratefully acknowledge support from the Nonclinical Biostatistics Scholarship awarded by the Nonclinical Biostatistics Working Group of the Biopharmaceutical Section of the American Statistical Association.}\hspace{.2cm}\\
    Department of Statistics\\
    University of California, Santa Cruz\\
    and \\
    Tony Pourmohamad \\
    Data and Statistical Sciences\\
    Genentech, Inc.\\
    and \\
    Bruno Sans\'{o}\\
    Department of Statistics\\
    University of California, Santa Cruz}
  \maketitle
} \fi

\if1\blind
{
  \bigskip
  \bigskip
  \bigskip
  \begin{center}
    {\LARGE\bf Tolerance Intervals Using Dirichlet Processes}
\end{center}
  \medskip
} \fi

\bigskip
\begin{abstract}
In nonclinical pharmaceutical development, tolerance intervals are critical in ensuring product and process quality. They are statistical intervals designed to contain a specified proportion of the population with a given confidence level. Parametric and non-parametric methods have been developed to obtain tolerance intervals. The former work with small samples but can be affected by distribution misspecification. The latter offer larger flexibility but require large sample sizes. As an alternative, we propose Dirichlet process-based Bayesian nonparametric tolerance intervals to overcome the limitations. We develop a computationally efficient tolerance interval construction algorithm based on the analytically tractable quantile process of the Dirichlet process. Simulation studies show that our new approach is very robust to distributional assumptions and performs as efficiently as existing tolerance interval methods. To illustrate how the model works in practice, we apply our method to the tolerance interval estimation for potency data.
\end{abstract}

\noindent%
\vfill

\newpage
\spacingset{2} 

\section{Introduction}
\label{sec:intro}
Tolerance intervals play a critical role in pharmaceutical and biopharmaceutical development by quantifying the range within which a specified proportion of product or process measurements are expected to fall. They provide direct evidence of process consistency \citep{hamada:2002, dong:2015, dong:2015b}, product stability \citep{montes2019simple,schwenke:2021, oliva:2025}, and assay reproducibility \citep{tendong2020ac50}, key attributes required for demonstrating control of critical quality characteristics and ensuring patient safety \citep{ichq6a1999, fda2011process}. In early research and development, tolerance intervals can help identify process variability and assess comparability across manufacturing campaigns \citep{lewis:2022}; in later stages, they serve as formal quality benchmarks for regulatory decision making \citep{little2016tolerance}.

Classical tolerance interval methodologies can be broadly classified into parametric and nonparametric approaches \citep{krishnamoorthy2009statistical, hahn:meeker:2017}. Parametric methods, which assume an underlying distributional form such as normality, are appealing for small samples and tractable inference when model assumptions hold \citep{wilks1941determination, howe1969}. However, they can be highly sensitive to distributional misspecification, leading to intervals that under- or over-cover the true population proportion \citep{krishnamoorthy2006improved}. Nonparametric approaches, by contrast, make minimal assumptions about the data-generating process and provide robust coverage guarantees \citep{hahn1970simple, young2014nonparametric}, but often at the cost of efficiency, especially in small-sample or high-heterogeneity settings common in pharmaceutical development \citep{cho:2021}.

Recent advances in Bayesian nonparametric modeling offer a powerful alternative framework for constructing tolerance intervals. By leveraging flexible priors such as the Dirichlet process (DP) \citep{ferguson1973bayesian, ferguson1974prior} and its hierarchical extension via mixture of Dirichlet process (MDP) \citep{antoniak1974mixtures}, it becomes possible to model complex, multimodal, or heavy-tailed distributions. This hierarchical structure is particularly attractive to robusify assumptions about distributional families in preclinical or multi-batch contexts, where data may exhibit unique distributional features.

Within this framework, tolerance intervals can be expressed in two complementary Bayesian formulations \citep{guttman1970statistical, hamada2004bayesian, chen2022bayesian}: 
\begin{enumerate}
\item $(\beta,\gamma)$-tolerance intervals, which ensure that a specified proportion $\beta$ of the population is covered with at least probability $\gamma$; and 
\item $\beta$-expectation tolerance intervals, which guarantee the same population coverage on average with respect to the posterior distribution. 
\end{enumerate}

These two formulations reflect distinct but related inferential goals, posterior coverage probability versus posterior expected coverage, and together capture the major decision-theoretic perspectives encountered in pharmaceutical quality assessment.

In this paper, we propose a Bayesian nonparametric framework for tolerance interval construction based on the DP prior. This framework leverages the analytical tractability of the DP posterior quantile process to provide flexible, computationally efficient, and distributionally robust interval estimates \citep{hjort2007nonparametric}. Because the method models the entire sampling distribution rather than fixed parameters, it remains valid across a wide variety of data-generating processes. This is particularly valid for the expectation tolerance intervals proposed in this paper, that are focused on expected coverage, and thus  are less sensitive to the tail behavior of the data distribution. 

The remainder of the paper is organized as follows. Section~\ref{sec:methods} reviews the concept of Bayesian tolerance intervals, the Dirichlet process prior, and quantile processes, establishing the foundation for our proposed framework. Section~\ref{sec:bnpti} introduces the construction of one-sided and two-sided tolerance intervals under the Dirichlet process model, highlighting both the probability and expectation formulations that arise from different decision-layer loss functions. Section~\ref{sec:simulations} presents simulation studies, and Section \ref{sec:potency} presents a real-data example taken from biopharmaceutical drug development. Both sections illustrate the performance and interpretability of the proposed methods. Finally, Section~\ref{sec:discussion} concludes with a discussion of extensions, limitations, and directions for future research.

\section{Methodology}
\label{sec:methods}

\subsection{Tolerance Intervals}
\label{subsec:ti}
Let $\bm{X} = (X_1,X_2,...,X_n)$ be an independent and identically distributed ($i.i.d.$) continuous random sample, where each $X_i$ has cumulative distribution function (cdf) $F_X(\theta)$. Further, let $Y$ be a random variable such that $Y\sim F_X(\theta)$, representing future unobserved data drawn from the same distribution as the $X_i$'s. There are several types of statistical tolerance regions used to quantify uncertainty about population coverage. Two of the most common are $(\beta,\gamma)$-tolerance intervals, which control coverage with specified probability, and $\beta$-expectation tolerance intervals, which control coverage on average with respect to the distribution of $Y$. Both play important roles in pharmaceutical and quality-control applications. To build intuition, we first review the $(\beta,\gamma)$ definition, as it provides a convenient link between frequentist and Bayesian formulations, and later (Section \ref{subsec:betaexpectation}) show how analogous reasoning leads to $\beta$-expectation intervals. In general, we say that the set $\mathcal{R}(\bm{X})$ is a $\beta$-content tolerance region at confidence level $\gamma$ if 
\begin{align}\label{toleranceregion}
\text{Pr}(P_X(\mathcal{R}(\bm{X}))\geq \beta)\geq \gamma,
\end{align}
where $P_X(\mathcal{R}(\bm{X})) = \text{Pr}(Y\in \mathcal{R}(\bm{X})) = \mathcal{C}(\bm{X})$ represents the proportion of the population covered by $\mathcal{R}(\bm{X})$. Thus, $\mathcal{C}(\bm{X})$ is often referred to as the coverage of the region $\mathcal{R}(\bm{X})$. As pointed out in \cite{chen2022bayesian}, $\mathcal{R}(\bm{X})$ is a {\em frequentist} $(\beta,\gamma)$-tolerance region if and only if the coverage $\mathcal{C}(\bm{X})$ is a confidence set for $\theta$ of confidence level $\gamma$. Now, the definition in (\ref{toleranceregion}) is easily extendable to the Bayesian paradigm by working with the posterior distribution of $\theta|\bm{X}$ \citep[see][]{aitchison1964two}. Under the Bayesian paradigm, $\mathcal{R}(\bm{X})$ is the $\beta$-content tolerance region at confidence level $\gamma$ if
\begin{align}\label{bayestoleranceregion}
\text{Pr}(P_X(\mathcal{R}(\bm{X}))\geq \beta|\bm{X})\geq \gamma, 
\end{align}
where we note that the probability in (\ref{toleranceregion}) is now computed as a posterior probability in (\ref{bayestoleranceregion}). Similarly, $\mathcal{R}(\bm{X})$ is a Bayesian $(\beta,\gamma)$-tolerance region if and only if the coverage $\mathcal{C}(\bm{X})$ is a credible set at level $\gamma$. An alternative Bayesian formulation replaces the probability constraint in (\ref{bayestoleranceregion}) with an expectation over the posterior, leading to $\beta$-expectation tolerance intervals discussed in  Section \ref{subsec:betaexpectation}.

To develop the ideas concretely, we first focus on the construction of $(\beta,\gamma)$-tolerance intervals. We make the simplifying assumption that $\bm{X}$ is a continuous univariate random sample, and thus $\mathcal{R}(\bm{X}) \subset \mathbb{R}$ is an interval on the real line. In particular, $\mathcal{R}(\bm{X})$ is an interval of the form $[L,\infty)$, $(-\infty,U]$, or [L,U], and is denoted as a one-sided lower, one-sided upper, or two-sided $(\beta,\gamma)$-tolerance interval, respectively. Finding the value of $L$, $U$, or both, is a non-trivial task and requires satisfying either (\ref{toleranceregion}) or (\ref{bayestoleranceregion}), depending on the chosen statistical paradigm. Regardless of the paradigm, the coverage simplifies as follows: 
\begin{itemize}
    \item For a one-sided lower interval $[L,\infty)$, the coverage is $\mathcal{C}(\bm{X}) = 1-F_X(L;\theta)$,
    \item For a one-sided upper interval $(-\infty,U]$, the coverage is $\mathcal{C}(\bm{X}) = F_X(U;\theta)$,
    \item For a two-sided interval $[L,U]$, the coverage is $\mathcal{C}(\bm{X}) = F_X(U;\theta) - F_X(L;\theta)$.
\end{itemize}
In the case of two-sided $(\beta,\gamma)$-tolerance intervals, the values of $L$ and $U$ must be determined jointly. However, direct analytical solutions are often unavailable, necessitating the use of approximations or optimization routines to ensure the interval achieves the desired coverage. In contrast, computing the limit ($L$ or $U$) for a one-sided interval is typically more straightforward, as it only requires solving for a single boundary to satisfy the coverage condition. A simple way to construct a two-sided tolerance interval is to intersect two one-sided tolerance intervals, each with a credible or confidence level of $\gamma$. However, the intersection of two one-sided intervals does not necessarily yield the shortest possible interval that satisfies the desired coverage. For this reason, specialized methods are often required to construct two-sided intervals that are both optimal in length and satisfy the coverage criterion.

\subsection{Bayesian Nonparametrics}
\label{subsec:bnp}
To construct tolerance intervals under minimal distributional assumptions, we consider a Bayesian nonparametric (BNP) approach based on the Dirichlet process (DP). BNP inference provides a framework for modeling probability distributions with infinitely many parameters, allowing practitioners to flexibly model infinite-dimensional objects such as distribution functions, density functions, or regression functions. This flexibility reduces reliance on restrictive assumptions, improving robustness to model misspecifications. Readers interested in a further introduction to BNP methods are encouraged to see \cite{walker1999bayesian}, \cite{muller2004nonparametric}, \cite{hanson2005bayesian}, and \cite{muller2013bayesian}.

Tolerance interval estimation based on parametric distributional assumptions can suffer from misspecification when the chosen model poorly reflects the true data-generating process. To address this limitation, we propose a new procedure for constructing tolerance intervals using Bayesian nonparametric priors, allowing for greater flexibility by modeling distributions over a wide range of function spaces. This approach estimates the distribution in a fully nonparametric way while simultaneously constructing tolerance intervals.

The Dirichlet Process (DP), introduced by \cite{ferguson1973bayesian} and \cite{ferguson1974prior}, is a popular BNP prior for density estimation. A random probability measure $F$ follows a DP with concentration parameter $a \in\mathbb{R}_+$ and base measure $F_0$ if and only if, for any measurable finite partition $\{A_l\}_{l=1}^L$ of the sample space of $F$, the vector of probabilities $(F(A_1), F(A_2), ..., F(A_L))$ follows the Dirichlet distribution with parameters\\ $(a F_0(A_1), a F_0(A_2), ..., a F_0(A_L))$. A key reason the DP is widely used in BNP density estimation is its conjugacy with $i.i.d.$ random samples. For example, under the model
\begin{equation}
\label{eq:model_toy}
    X_i|F \stackrel{iid}{\sim} F \text{ for } i=1,2,...,n, \,\,\,\text{ and }\,\,\,
    F \sim \text{DP}(a, F_0),
\end{equation}
then by Bayes' theorem we have that
\begin{equation}\label{eq:dp_post}
    F|\bm{X} \sim \text{DP}\left(a+n, \frac{a}{a+n}F_0 + \frac{n}{a+n}F_n\right)
\end{equation}
where $F_n$ is the empirical distribution function of $\bm{X}$. The posterior distribution $F|\bm{X}$ follows a DP again, with updated parameters. The posterior distribution is a convex combination of the prior center measure $F_0$ and the empirical distribution $F_n$, weighted by $a$ and $n$. As $a \rightarrow 0$, the posterior tends to $F_n$, while as $a \rightarrow \infty$, it tends to $F_0$. Thus, the posterior offers a compromise between prior information and the observed data, with the concentration parameter $a$ controlling the balance. Furthermore, the analytic form of the posterior significantly reduces computational complexity in practical applications.

\subsection{Quantile Processes}
\label{subsec:qp}
Since tolerance limits correspond to specific quantiles of the underlying distribution, the DP quantile process provides a natural analytic foundation for interval construction. In addition to \eqref{eq:model_toy}, let $Y|F\sim F$, conditionally independent of the other $X_i$ given $F$. In this context, the construction of a tolerance interval for $Y$ greatly benefits from the structure provided by the DP in \eqref{eq:model_toy}. Specifically, the quantile process $Q(q)=\inf\{t:F(t)\geq q\}, q\in[0,1],$ has been extensively studied by \cite{hjort2007nonparametric}. Their work provides a significant tool for deriving tolerance intervals using the DP prior. From their results, the distribution of $Q(q)$ under the DP can be expressed analytically.  This avoids the need for computationally intensive numerical algorithms when constructing tolerance intervals. The distribution function of $Q(q)$, denoted by $H_{0,a,q}$, is given in \cite{hjort2007nonparametric} as
\begin{equation}
    \begin{split}
    H_{0,a,q}(x) &= P_Q(Q(q)\leq x) = P_F(F(x) \geq q) \\
    &= 1 - \text{Be}(q; a F_0(x), a (1-F_0(x))) 
    = \text{Be}(1-q; a (1-F_0(x)), a F_0(x))
\end{split}
\label{eq:hjort_dpquant1pt}
\end{equation}
where $\text{Be}(q_0;a_0,b_0)$ is the cdf of the Beta distribution at the point $q_0$ with parameters $a_o$ and $b_0$. This result follows directly from the definition of the DP and the relationship between the Beta and Dirichlet distributions. The posterior quantile process for $Q(q)$, updated with observations $X$, takes a similar form: 
\begin{equation}
    \begin{split}
    H_{n,a,q}(x) &= P_{Q}(Q(q)\leq x |\bm{X}) = P_{F}(F(x) \geq q |\bm{X}) \\
    &= 1 - \text{Be}(q; (a F_0+n F_n)(x), (a (1-F_0)+n (1-F_n))(x)).
\end{split}
\label{eq:hjort_dpquant1pt}
\end{equation}
Note that the last expression is written as $Be(1-q; a (1-F_0)(x) + n - i, a F_0(x) + i)$ if $x$ is such that $X_{(i)} \leq x < X_{(i+1)}$. This result holds even if $F|\bm{X}$ has countably many jumps.

\section{Bayesian Nonparametric Tolerance Intervals}
\label{sec:bnpti}
Connecting the methodological pieces of Sections \ref{subsec:ti}--\ref{subsec:qp}, we develop a novel strategy for constructing one- and two-sided tolerance intervals via quantile inference under a BNP framework. 

\subsection{Dirichlet Process Based Tolerance Intervals}
Starting with the construction of the simpler one-sided tolerance intervals, the Bayesian nonparametric one-sided DP-based $(\beta, \gamma)$-tolerance interval for $Y$ can be derived from the posterior quantile process $H_{n,\alpha,\beta}$ defined in \eqref{eq:hjort_dpquant1pt}. More specifically, to obtain a one-sided tolerance interval of the form $(-\infty, L(\bm{X})]$, we find $L(\bm{X})$ such that $H_{n,a,\beta}(L(\bm{X})) = \gamma$. Analogously, for the upper one-sided tolerance interval $[U(X), \infty)$, we find $U(\bm{X})$ such that $H_{n,a,1-\beta}(U(\bm{X})) = 1-\gamma$.
Here, any root-finding algorithms can be used to solve for $L(\bm{X})$ or $U(\bm{X})$ (e.g., Brent’s method \citep{brent1973algorithms} or bisection \citep{press2007numerical}). Even a simple grid search using a loop can be used to solve the equation because $H_{n,a,\beta}$ is monotonic in $x$ with $n$ (finite) number of jumps.

Extending this idea to two-sided tolerance intervals is a more challenging task. However, a simple approach is to construct a two-sided tolerance interval by intersecting two one-sided tolerance intervals with a credible level of $1-(1-\delta)/2$. By Bonferroni's theorem \citep{casella2002statistical}, the resulting interval maintains the desired coverage level of $\gamma$. The simplicity of this approach is a key advantage. However, this method is often criticized for being overly conservative, potentially leading to unnecessarily wide intervals.

To better understand how tolerance limits behave in the DP model, it is instructive to first consider the simpler case of a one-sided interval. Defining tolerance limits directly within the decision layer allows us to track how the limits vary with $a$ and $n$. To illustrate the behaviors, we focus on a one-sided tolerance interval with its upper limit $U(\bm{X})$.
Recalling \eqref{eq:dp_post} and \eqref{eq:hjort_dpquant1pt}, when $a\rightarrow 0$,
\begin{equation}
\begin{split}
\gamma &= H_{n,0,\beta}(U(\bm{X})) 
= 1-\text{Be}(\beta; n F_n(U(\bm{X})), n - n F_n(U(\bm{X}))),
\end{split}
\end{equation}
where $\text{Be}(q_0;a_0,b_0)$ is the cdf of the Beta distribution at the point $q_0$ with parameters $a_0$ and $b_0$ again. We can then choose $X_{(m)}\leq U(\bm{X}) < X_{(m+1)}$, or simply $U(\bm{X})=X_{(m)}$, where $X_{(i)}$ is the $i$-th order statistic of $\bm{X}$, and $m$ is the smallest integer satisfying 
\begin{equation}
1-\text{Be}(p; m, n-m) \geq \gamma.
\end{equation}
Equivalently, $m$ can be found as the smallest integer such that $P(W\leq m;n+1,\beta)\geq \gamma$ where $W\sim \text{Bin}(n+1,\beta)$, using the known Beta-Binomial relationship: $P(W\geq k;n,p)=P(U\leq p)$ when $U\sim \text{Beta}(k, n-k+1)$ and $W\sim \text{Bin}(n,p)$.

This result is very similar to the frequentist nonparametric tolerance interval based on Wilks' theorem and the distribution of order statistics. In the frequentist case, the nonparametric $(\beta, \gamma)$ tolerance upper limit is chosen as $X_{(m)}$ where $m$ is the smallest integer satisfying
\begin{equation}
        \gamma \leq P_{X_{(m)}}[P_X(X_{(m)}\geq X | X_{(m)}) \geq \beta] 
        = P_{X_{(m)}}[F_X(X_{(m)})\geq \beta]
        = P(U_{(m)}\geq \beta) 
    \label{eq:freqnonparam_tol}
\end{equation}
where $U_{(m)}\sim \text{Beta}(m, n-m+1)$, or equivalently, $m$ is the smallest integer such that $P(W\leq m-1;n,p) \geq \gamma$, where $W\sim \text{Bin}(n,p)$.
Despite the different underlying reasoning, both the Bayesian DP-based and frequentist nonparametric procedures select the tolerance limit at one of the observed data points, and the calculation of 
$m$ involves Beta distributions differing by just 1 in the second parameter. Heuristically, this can be interpreted as the Bayesian limit ``losing" roughly the information of one additional observation compared to the frequentist limit, because the DP posterior considers $n+1$ partitions given $n$ observations when $a\rightarrow0$. This difference becomes negligible as $n$ increases, so the DP-based nonparametric Bayesian tolerance limit performs comparably to the frequentist nonparametric limit. As a result, in most practical cases, the two procedures select the same limit for a given $\gamma$ and $\beta$. For example, when $n = 100$, $\gamma=0.95$, and $\beta = 0.95$, both methods select the second largest observation ($m=99)$ as the upper tolerance limit. For $n = 1000$ with the same $\gamma$ and $\beta$, the selected $m$ is 962, so the 962nd observation is the upper limit in both approaches.

This connection allows us to leverage known properties from the frequentist nonparametric tolerance interval literature in the Bayesian setting. For instance, we can adapt rules for sample size. In the frequentist approach, the minimum $n$ such that $n \geq \log(1-\gamma)/\log(\beta)$ ensures a $(\beta, \gamma)$ upper tolerance limit can be constructed; otherwise, the limit is simply the maximum observed value, since no valid $m$ exists in \eqref{eq:freqnonparam_tol}. Extending this reasoning, a rough rule of thumb in the Bayesian case is $n \geq \log(1-\gamma)/\log(\beta) + 1$ for constructing a $(\beta, \gamma)$ limit when $a$ is very close to zero. For example, with $\gamma = \beta = 0.95$, this yields $n \geq 60$. Fortunately, this is not a strict requirement, because the prior distribution provides additional information unless $a = 0$.

On the other hand, if $a \to \infty$ while $n$ is fixed, the posterior $F \mid \bm{X}$ is dominated by the prior, $F \mid \bm{X} \approx F_0.$ In this case, the tolerance limit is simply the $\beta$- or $(1-\beta)$-quantile of the prior $F_0$. Similarly, if $n \to \infty$ with $a$ fixed, the posterior is dominated by the empirical distribution, $F \mid \bm{X} \approx F_n$, which converges to the true data-generating distribution at rate $\sqrt{n}$ provided the support of $F_0$ is not too restrictive. Consequently, the tolerance limit converges to the correct value asymptotically. These asymptotic properties guide practical prior elicitation. The parameter $a$ can be interpreted as the effective sample size contributed by the prior. Choosing $F_0$ to represent a reasonable ``average" distribution ensures that posterior inference adapts appropriately as data accumulate.
\color{black}

\subsection{Hierarchical Extension}\label{subsec:hierarchical}
The analysis above demonstrates that the two parameters of the DP, $a$ and $F_0$, significantly influence the resulting tolerance interval, especially when the sample size is small. However, specifying appropriate values for these parameters can be challenging when prior information is limited. To address this, we can robustify our DP model by placing priors over $a$ and $F_0$, allowing Bayes’ rule to automatically determine these parameters in the posterior distribution.
 
A natural extension of the DP framework is through a mixture of Dirichlet processes, as introduced by \cite{antoniak1974mixtures}. Consider the following hierarchical model:
\begin{equation}
        X_i|F \stackrel{iid}{\sim} F \text{ for } i=1,2,...,n;
        \;\;\;F|\xi,a \sim \text{DP}(a, F_0(\xi))
\end{equation}
for $\xi \sim p(\xi)$ and $a \sim \text{Gamma}(a_a, b_a)$ where $F_0$ is a distribution parameterized by $\xi$ and the hyperparameters $a_a$ and $b_a$ are given.
The Gamma prior for $a$ is chosen for computational convenience during posterior updates. After observing data, posterior inference can be performed using a generalized P\'{o}lya urn scheme by integrating $F$ out of the model. Let $k$ denote the number of distinct values in $\{X_i\}^{n}_{i=1}$, and let $S$ represent a partition structure of the indices $i=1,2,...,n$, where $i$ and $i^\prime$ belong to the same cell if $X_i=X_{i^\prime}$. The joint posterior distribution of $(a, S, \xi)$ given the data is proportional to $p(X_1, X_2, ..., X_n|F_0(\xi), S)p(S|a)p(a)p(\xi)$.
This expression separates the components for $a$ and $\xi$ because the partition structure of the DP depends on $a$ only through $k$, and both $S$ and $k$ are observable from the data. Introducing an auxiliary variable $\eta$, the full conditional posterior distributions are
\begin{equation} \label{eq:mdp_post_fullcond}
    \begin{split}
        p(\xi|...) &\propto p(\xi)\left(\prod_{i=1}^n f_0(X_i;\xi)\right)\left(\prod_{i=1}^n \frac{n_i}{a+i-1}\right)\\
        p(a|...) &\sim \text{Gamma}(a_a+k, b_a - \log{\eta})\\
        p(\eta|...) &\sim \text{Beta}(a, n),
    \end{split}
\end{equation}
where $n_i$ is the number of ties with $X_i$ before the $i^{th}$ data point, and $f_0$ is the density function of $F_0$.
For more details, see chapter 4.5 of \cite{ghosal2017fundamentals}.
Note that $k=n$ almost surely, and all $n_i=1$ if $X_i$ comes from a continuous distribution.

Using the posterior samples of $(a, \xi)$, we can obtain posterior samples at specific quantiles for the quantile process of $F|X_1,X_2,...,X_n,a,\xi \sim \text{DP}(a+n, \frac{a}{a+n}F_0(\xi)+\frac{n}{a+n}F_n)$ by \eqref{eq:hjort_dpquant1pt}. These posterior samples enable the construction of a tolerance interval by controlling $\gamma$ over the samples.

\subsection{Expectation tolerance intervals}\label{subsec:betaexpectation}
Another type of Bayesian tolerance interval is the \emph{expectation tolerance interval}. 
We define a Bayesian $\beta$-expectation tolerance interval $[L^*(\bm{X}),U^*(\bm{X})]$ as any interval satisfying
\begin{equation}
\int P\big[X\in[L^*(\bm{X}),U^*(\bm{X})] \mid \bm{X}, \theta \big]\, p(\theta \mid \bm{X})\, d\theta 
= E_{\theta\mid \bm{X}}\big[P\big[X\in[L^*(\bm{X}),U^*(\bm{X})] \mid \bm{X},\theta \big]\big] \geq \beta,
\label{eq:beta_expectation_interval}
\end{equation}
where the expectation is taken with respect to the posterior distribution of $\theta$, not the data $\bm{X}$. Under the common assumption that a new observation $Y$ is conditionally independent of the observed data $\bm{X}$ given $\theta$, \eqref{eq:beta_expectation_interval} reduces to
\begin{equation}
E_{\theta\mid \bm{X}}\big[P\big[Y\in[L^*(\bm{X}),U^*(\bm{X})] \mid \theta \big]\big] \geq \beta.
\end{equation}

In other words, the interval $[L^*(\bm{X}),U^*(\bm{X})]$ is expected to cover a new observation with probability at least $\beta$ on average with respect to the posterior distribution of $\theta$. 
The key distinction from $(\beta, \gamma)$-tolerance intervals is that we do not directly control the chance of coverage in each posterior realization; rather, we control it on average across the posterior.

\cite{chen2022bayesian} showed that these two types of tolerance intervals correspond to different loss functions at the decision layer. That is, these two types of Bayesian tolerance intervals represent alternative choices for summarizing the posterior distribution of a stochastic prior-likelihood model to estimate the underlying density.
Building on this idea, we introduce a decision-level method for constructing expectation tolerance intervals under the same stochastic DP model for $X_i$ and $F$.

Among many possibilities, a natural decision rule is as follows. Choose $q_1, q_2 \in [0,1]$ such that $q_1 < q_2$ and $|q_2-q_1| = \beta$, and then set
\begin{align}
[L^*(\bm{X}), U^*(\bm{X})] = [E[Q(q_1) \mid \bm{X}],\, E[Q(q_2) \mid \bm{X}]],
\end{align}
where $Q(q)$ is the posterior quantile process of $F \mid \bm{X}$. Compared to $(\beta, \gamma)$-tolerance intervals, we take posterior means rather than posterior $\gamma$ or $1-\gamma$ quantiles. 
Consequently, these limits are generally not located at observed data points, even in the limiting case $\alpha \to 0$, and they vary continuously with $q_1$ and $q_2$. 
In particular, they are not the same as the $(q_1, q_2)$-quantiles of the weighted posterior $\frac{a}{a+n} F_0 + \frac{n}{a+n} F_n$.

For practical computation, the posterior expected value of $Q(q)$ can be calculated using the tail-sum formula of positive random variables \citep{hjort2007nonparametric}:
\begin{equation}
\begin{split}
E[Q(q) \mid \bm{X}] 
&= \int_0^\infty P[Q(q)\geq x \mid \bm{X}] \, dx - \int_{-\infty}^0 P[Q(q)\leq x \mid \bm{X}] \, dx \\
&= \int_{0}^{\infty} \text{Be}(q; aF_0(x)+nF_n(x), a(1-F_0(x))+n(1-F_n(x)))\, dx \\
&\quad - \int_{-\infty}^{0} \text{Be}(1-q; a(1-F_0(x))+n(1-F_n(x)), aF_0(x)+nF_n(x))\, dx,
\end{split}    
\label{eq:expectation_Q}
\end{equation}
where $\text{Be}(\cdot;\cdot,\cdot)$ denotes the Beta CDF. 
For $a>0$, the integrands are monotone in $x$ with finitely many jumps and vanish as $x$ moves far from the data range. Thus, a simple grid-based evaluation suffices for numerical integration. Equation \eqref{eq:expectation_Q} clearly reveals the continuity of our estimate of the quantile function on $q$. Thus, even though the Dirchlet process produces discrete distributions, we have obtained a continuous estimate of the quantile function of the data generating process. In the special case $a \to 0$, \eqref{eq:expectation_Q} reduces to
\begin{equation}
E[Q(q) \mid \bm{X}] = \sum_{i=1}^n {n-1 \choose i-1} q^{i-1} (1-q)^{n-i} x_{(i)},
\end{equation}
with $E[Q(0)\mid \bm{X}] = x_{(1)}$ and $E[Q(1)\mid \bm{X}] = x_{(n)}$. Finally, because this decision-making procedure depends only on the marginal posterior of the DP layer, it can be applied to any DP-based density estimation model, including hierarchical extensions such as the mixture of Dirichlet processes described in Section~\ref{subsec:hierarchical}.

\section{Simulation Study}
\label{sec:simulations}

\subsection{$(\beta,\gamma)$-tolerance intervals}
To evaluate the performance of the proposed $(\beta,\gamma)$-tolerance intervals, we conducted a simulation study. Two data-generating processes (DGPs) were considered: $N(0,2^2)$ and $\text{Laplace}(0,2)$ as a thicker tail alternative to the Gaussian case. For each DGP, we simulated i.i.d. samples of size $n$ and fitted one-sided lower tolerance intervals with parameters $(\beta=0.95, \gamma=0.95)$. 
Sample sizes $n=10, 30, 50, 100$, and $1000$ were considered, and each setting was repeated $K=10{,}000$ times. 
The goal was to monitor how well the procedure matches the nominal content $\beta=0.95$ with coverage $\gamma=0.95$. Specifically, let $\bm{X}_k = (X_{k1}, X_{k2}, \dots, X_{kn})$, $k=1, \dots, K$, denote the simulated datasets, and let $\{[L(\bm{X}_k),U(\bm{X}_k)]\}_{k=1}^K$ be the corresponding fitted intervals, with $L(\bm{X}_k)=-\infty$. 
The coverage probability of each interval was evaluated as 
\begin{align}
    CP([L(\bm{X}_k),U(\bm{X}_k)],F)=F(U(\bm{X}_k))-F(L(\bm{X}_k)),
\end{align}
and the mean coverage and undercoverage rates were computed as
\begin{align}
  \text{mean coverage}(F) = \frac{1}{K}\sum_{k=1}^K CP([L(\bm{X}_k),U(\bm{X}_k)],F)  
\end{align}
and
\begin{align}
\text{undercoverage rate}(F) = \frac{1}{K} \sum_{k=1}^K I(CP([L(\bm{X}_k),U(\bm{X}_k)],F)<\beta),
\end{align}
with $I(\cdot)$ the usual indicator function. The mean coverage should be close to $\beta$, and the undercoverage rate should approximate $1-\gamma$ if the procedure performs well. In addition, we monitor the mean upper limit as an analog to evaluate interval lengths. A shorter interval (a smaller mean upper limit) matching $\beta$ and $\gamma$ is preferred.

Table \ref{tab:sim_bayes_onesided_normal} summarizes the results under the $N(0,2^2)$ DGP.
For comparison, we included the Bayesian parametric method under a Gaussian likelihood with a noninformative prior (equivalent to the frequentist parametric approach) and the frequentist nonparametric method. When the parametric assumption is correct, these referenced approaches perform nearly perfectly even for small $n$. Compared to this benchmark, the DP-based Bayesian nonparametric method with a weak prior ($a=10, F_0 = N(0,2^2)$) is slightly less efficient for small $n$, but performance improves as $n$ increases or if the prior is stronger (e.g., $a=100, F_0 = N(0,2^2)$). 
This trade-off reflects the flexibility of the nonparametric approach. 

Notably, the method is robust: even when the prior $F_0$ is misspecified (e.g., $t_5$ scaled by 2 or Laplace with scale 2), the posterior tolerance limits converge to the correct values as $n$ grows. Thus, moderate sample sizes ($n=20$--$30$) are sufficient for reliable performance. Additionally, the DP-based method requires less data than frequentist nonparametric approaches, as prior information helps construct tolerance limits and quantify uncertainty even for small $n$. The MDP setting, with $a\sim \text{Gamma}(1,1)$ and $F_0 = N(\mu,\sigma^2)$ with vague priors $\mu\sim N(0,10^2)$ and $\sigma^2 \sim \text{Inv-Gamma}(1,1)$, also performs well, even for very small $n$, though it entails an increased computational burden.

A notable feature of DP-based tolerance limits is their non-linear improvement with $n$. 
As Table~\ref{tab:sim_bayes_onesided_diff_n} shows, the limits occasionally jump sharply toward the correct value and then improve gradually until the next jump. This behavior arises from the discreteness of the DP posterior; estimated quantiles move when an additional data point substantially affects the posterior in the relevant tail. Additional data provide incremental support between jumps, stabilizing the estimate. 

Table~\ref{tab:sim_bayes_onesided_laplace} reports results for the $\text{Laplace}(0,2)$ DGP. 
Parametric tolerance interval construction is challenging under the Laplace DGP, but the DP-based approach alleviates this difficulty. Even with moderate prior strength, DP-based intervals perform well across different choices of $F_0$. Stronger prior information (larger $a$) improves performance, as illustrated by DP$(a=100, F_0 = \text{Laplace}(0,2))$. The MDP also yields reasonable results under vague priors, demonstrating robustness.

In summary, the proposed Bayesian DP- and MDP-based methods are: 1) robust to prior misspecification and the true underlying DGP, 2) perform well for moderate $n$ without requiring large prior strength, 3) are applicable to asymmetric or truncated distributions, 4) can be used for any distribution, including those without available parametric tolerance interval routines.

\clearpage
\begin{table}
\centering
\resizebox{0.7\textwidth}{!}{
\begin{tabular}{l l | c c c c c} 
Method &  & n=10 & n=30 & n=50 & n=100 & n=1000 \\
 \hline
Normal (noninformative)  & undercoverage & 0.0475 & 0.0502 & 0.0504 & 0.0484 & 0.0502 \\
 & mean coverage  & 0.9894 & 0.9813 & 0.9768 & 0.9710 & 0.9578 \\
 & mean upper limit & 5.674 & 4.4002 & 4.1067 & 3.8448 & 3.4550 \\
Frequentist Nonparametric
 & undercoverage & -- & -- & -- & 0.0363 & 0.0466 \\
 & mean coverage  & --  & -- & -- & 0.9801 & 0.9611 \\
 & mean upper limit & -- & -- & -- & 4.2951 & 3.5367 \\
\hline
DP$(a=10,F_0 = N(0,2^2))$
 & undercoverage  & 0 & 0.2142 & 0.0754 & 0.1175 & 0.0584 \\
 & mean coverage  & 0.9826 & 0.9735 & 0.9803 & 0.9712 & 0.9600 \\
 & mean upper limit & 4.2794 & 4.1642 & 4.4508 & 3.9041 & 3.5112 \\
DP$(a=100,F_0 = N(0,2^2))$
 & undercoverage & 0 & 0 & 0 & 0.0059 & 0.0452 \\
 & mean coverage  & 0.9768 & 0.9747 & 0.9730 & 0.9704 & 0.9599 \\
 & mean upper limit & 3.9968 & 3.9345 & 3.8835 & 3.8031 & 3.5070 \\
DP$(a=10,F_0 = \text{Laplace}(0,2))$
 & undercoverage & 0 & 0 & 0.0818 & 0.0361 & 0.0452 \\
 & mean coverage  & 0.9995 & 0.9893 & 0.9810 & 0.9791 & 0.9609 \\
 & mean upper limit & 6.5358 & 4.7046 & 4.5003 & 4.1847 & 3.5300 \\
DP$(a=100,F_0 = \text{Laplace}(0,2))$
 & undercoverage & 0 & 0 & 0 & 0 & 0.0099 \\
 & mean coverage  & 0.9986 & 0.9972 & 0.9953 & 0.9899 & 0.9641 \\
 & mean upper limit & 5.9930 & 5.5526 & 5.2168 & 4.6870 & 3.6108 \\
DP$(a=10,F_0 = 2t_5)$ & undercoverage & 0 & 0 & 0.0818 & 0.0361 & 0.0601 \\
 & mean coverage  & 0.9975 & 0.9823 & 0.9804 & 0.9761 & 0.9602 \\
 & mean upper limit & 5.6114 & 4.3879 & 4.4891 & 4.0424 & 3.5155 \\
DP$(a=100,F_0 = 2t_5)$
 & undercoverage & 0 & 0 & 0 & 0 & 0.0145 \\
 & mean coverage  & 0.9950 & 0.9921 & 0.9893 & 0.9836 & 0.9627 \\
 & mean upper limit & 5.1624 & 4.8519 & 4.6357 & 4.3089 & 3.5728 \\
\hline
MDP$(a\sim \text{Gamma}(1,1), F_0 = N(\mu, \sigma^2)$ 
 & undercoverage & 0.0970 & 0.1220 & 0.0750 & 0.0680 & 0.0680 \\
$\mu\sim N(0, 10^2)$ 
 & mean coverage  & 0.9837 & 0.9763 & 0.9787 & 0.9726 & 0.9587 \\
$\sigma^2\sim \text{Inv-Gamma}(1,1)$ 
 & mean upper limit & 5.6763 & 4.3093 & 4.275 & 3.9328 & 3.4789 \\
\hline
\hline
\end{tabular}
}
\caption{Simulation results for one-sided $(\beta=0.95, \gamma = 0.95)$ tolerance intervals under the $N(0, 2^2)$ DGP. Each entry summarizes 10,000 simulations. Frequentist nonparametric tolerance limits (Wilks' method) are undefined for $n<59$.}
\label{tab:sim_bayes_onesided_normal}
\end{table}

\begin{table}[]
    \centering
    \resizebox{0.9\textwidth}{!}{
    \begin{tabular}{l c|c c c c c}
         Method & n & undercoverage & mean coverage & 95\% CI & mean upper limit & 95\% CI \\
         \hline
          DP$(a = 1, F_0 = N(0,2^2))$ 
          & 57 & 0.0487 & 0.9833 & [0.9386 0.9996] & 4.6239 & [3.0857 6.6574] \\
          & 58 & 0.0448 & 0.9835 & [0.9403 0.9996] & 4.6310 & [3.114  6.6597] \\
          & 59 & 0.0457 & 0.9797 & [0.9401 0.9959] & {\bf 4.2167} & [3.1105 5.2950] \\
          & 60 & 0.0413 & 0.9720 & [0.9418 0.9958] & {\bf 3.9550} & [3.1407 5.2678] \\
          \color{blue}& 61 & 0.1797 & 0.9691 & [0.9315 0.9959] & 3.9061 & [2.9746 5.2824] \\
          \color{black}& 62 & 0.1685 & 0.9689 & [0.9135 0.9960] & 3.9133 & [2.7257 5.3011] \\
          & 63 & 0.1627 & 0.9692 & [0.9151 0.996 ] & 3.9236 & [2.7456 5.3070] \\
          & \vdots & \vdots& \vdots& \vdots& \vdots& \vdots\\
          & 90 & 0.0526 & 0.9782 & [0.9407 0.9972] & 4.2159 & [3.1208 5.5420] \\
          & 91 & 0.0532 & 0.9784 & [0.941  0.9972] & 4.2217 & [3.1268 5.5434] \\
          & 92 & 0.0476 & 0.9786 & [0.9417 0.9973] & 4.2307 & [3.1386 5.5524] \\
          & 93 & 0.0483 & 0.9789 & [0.9419 0.9973] & 4.2410 & [3.1425 5.5572] \\
          & 94 & 0.0436 & 0.9730 & [0.9427 0.9933] & {\bf 3.9381} & [3.156  4.9481] \\
          \color{blue} & 95 & 0.1399 & 0.9695 & [0.9372 0.9935] & 3.8612 & [3.0635 4.9681] \\
          \color{black} & 96 & 0.1336 & 0.9693 & [0.9276 0.9934] & 3.8626 & [2.9163 4.9563] \\
          & 97 & 0.1274 & 0.9696 & [0.927  0.9933] & 3.8699 & [2.9072 4.9494] \\
          \hline
    \end{tabular}}
\caption{Results from 10,000 simulations under a $N(0,2)$ DGP for the one-sided $(0.95,0.95)$ tolerance limit using a DP prior with $a=1$. The table reports empirical undercoverage, mean coverage with Monte Carlo intervals, and the mean upper tolerance limit. Bolded upper limits correspond to sample sizes at which the DP estimate exhibits an abrupt jump, reflecting the discrete nature of DP posterior updates in the upper tail.}
\label{tab:sim_bayes_onesided_diff_n}
\end{table}

\begin{table}
\centering
\resizebox{0.9\textwidth}{!}{
\begin{tabular}{l l | c c c c c} 
 Method &  & n=10 & n=30 & n=50 & n=100 & n=1000 \\
 \hline
Normal (noninformative) 
 & undercoverage  & 0.1126 & 0.1106 & 0.1096 & 0.1080 & 0.0774 \\
 & mean coverage  & 0.9786 & 0.9718 & 0.9687 & 0.9648 & 0.9561 \\
 & mean upper limit & 7.7998 & 6.1594 & 5.7686 & 5.4132 & 4.8792 \\
Frequentist Nonparametric
 & undercoverage  & -- & -- & -- & 0.0372 & 0.0464 \\
 & mean coverage  & -- & -- & -- & 0.9802 & 0.9609 \\
 & mean upper limit & -- & -- & -- & 6.9899 & 5.1210 \\
DP$(a=10, F_0=N(0, 2^2))$
 & undercoverage   & 0.5918 & 0.2086 & 0.0759 & 0.1161 & 0.0607 \\
 & mean coverage  & 0.9539 & 0.9694 & 0.9731 & 0.9704 & 0.9600 \\
 & mean upper limit & 5.2728 & 6.6613 & 6.1622 & 5.9949 & 5.0749 \\
DP$(a=10,F_0=\text{Laplace}(0, 2))$
 & undercoverage  & 0 & 0.2086 & 0.0759 & 0.1161 & 0.0607 \\
 & mean coverage  & 0.9827 & 0.9739 & 0.9806 & 0.9714 & 0.9600 \\
 & mean upper limit & 6.9120 & 6.7911 & 7.4867 & 6.0277 & 5.0749 \\
DP$(a=100,F_0=\text{Laplace}(0,2))$
 & undercoverage  & 0 & 0 & 0 & 0.006 & 0.0454 \\
 & mean coverage  & 0.9769 & 0.9748 & 0.9732 & 0.9704 & 0.9597 \\
 & mean upper limit & 6.1835 & 6.0490 & 5.9387 & 5.7399 & 5.0584 \\
MDP($a\sim\text{Gamma(1,1)},F_0 = N(\mu,\sigma^2))$ & undercoverage  & 0.1470 & 0.1380 & 0.0850 & 0.0780 & 0.0580\\
 $\mu\sim N(0,10^2)$& mean coverage  & 0.9761 & 0.97354 & 0.9749 & 0.9708 & 0.9583 \\
 $\sigma^2\sim\text{Inv-Gamma(1,1)}$& mean upper limit & 10.1999 & 6.8710 & 6.4114 & 5.8816 & 4.9854 \\
 \hline
 \hline
\end{tabular}}
\caption{Simulation results for one-sided $(\beta=0.95, \gamma = 0.95)$ tolerance intervals under the $\text{Laplace}(0, 2)$ DGP. Each entry summarizes 10,000 simulations. Frequentist nonparametric tolerance limits (Wilks' method) are undefined for $n<59$.}
\label{tab:sim_bayes_onesided_laplace}
\end{table}

\clearpage
\subsection{$\beta$-expectation tolerance intervals}
\subsubsection{Fully Bayesian $\beta$-expectation tolerance intervals}
For $\beta$-expectation intervals, we conduct a simulation study similar to the one in the previous section. Two DGPs are considered: $N(0,1)$ and $\text{Laplace}(0,1)$. For each DGP, we generate samples of size $n = 10, 30, 50, 100, 1000$ and fit two-sided $(\beta=0.95)$-expectation tolerance intervals, repeating the procedure 2,000 times. The mean coverages and interval lengths are then recorded.

Table \ref{tab:sim_bayes_expect_tol_N} summarizes results under the $N(0,1)$ DGP. Compared to $(\beta,\gamma)$ intervals with $\beta=0.95$ and $\gamma$ close to $1$, $\beta$-expectation intervals are shorter and exhibit less variability, since they are based on the expected value of the quantile process rather than its tail. These intervals generally achieve coverage close to the nominal $\beta$ on average, making them a practical choice unless stricter guaranteed coverage is required. 

Under small sample sizes, careful prior specification can help avoid undercoverage. In particular, a heavy-tailed prior mean measure $F_0$ with moderate precision $a$ (e.g., DP priors with $a=10, F_0=t_5$ or $a=10, F_0=\text{Laplace(0,1)}$) provides satisfactory expected tolerance limits. If strong prior information is available, choose a large value of the prior precision $\alpha$ to increase the influence of the prior distribution and improve the performance of the tolerance interval. For moderate sample sizes ($n \approx 30$), most prior settings yield acceptable results. As the sample size $n$ increases, all tolerance limits converge to the true quantiles regardless of prior choice.

The general convergence behavior observed for $(\beta,\gamma)$ intervals also applies to $\beta$-expectation intervals. In practice, this convergence is often faster and more apparent for $\beta$-expectation intervals, allowing the true quantiles to be estimated more reliably. These observations support a consistent approach to prior elicitation for both types of intervals. Table \ref{tab:sim_bayes_expect_tol_L} shows analogous results under the $\text{Laplace}(0,1)$ DGP. As expected, the method performs robustly across different DGPs, provided that the prior mean measure $F_0$ covers the full support of the true data-generating process.

\begin{table}[h]
\centering
\resizebox{0.9\textwidth}{!}{
\begin{tabular}{l l | c c c c c} 
 Method &  & n=10 & n=30 & n=50 & n=100 & n=1000 \\
\hline
DP$(a=0)$
& coverage & 0.7890 & 0.9028 & 0.9255 & 0.9400 & 0.9489 \\
& length & 2.8379 & 3.5406 & 3.7267 & 3.8548 & 3.9142 \\
DP$(a=10, F_0=N(0,1))$
& coverage & 0.9084 & 0.9236 & 0.9330 & 0.9417 & 0.9490 \\
& length & 3.4850 & 3.6857 & 3.7832 & 3.8666 & 3.9153 \\
DP$(a=10,F_0=t_5)$
& coverage & 0.9566 & 0.9466 & 0.9461 & 0.9476 & 0.9496 \\
& length & 4.1254 & 3.9957 & 3.9758 & 3.9616 & 3.9250 \\
DP$(a=10,F_0=\text{Laplace}(0,1))$
& coverage & 0.9733 & 0.9581 & 0.9535 & 0.9510 & 0.9499 \\
& length & 4.5038 & 4.1951 & 4.1013 & 4.022 & 3.9307 \\
DP$(a=100,F_0=N(0,1))$
& coverage & 0.9479 & 0.9474 & 0.9472 & 0.9473 & 0.9492 \\
& length & 3.8932 & 3.8951 & 3.8972 & 3.9014 & 3.9159 \\
DP$(a=100, F_0=t_5)$
& coverage & 0.9884 & 0.9845 & 0.9810 & 0.9740 & 0.9543 \\
& length & 5.0550 & 4.8564 & 4.7110 & 4.481 & 4.0070 \\
DP$(a=100,F_0=\text{Laplace}(0,1))$
& coverage & 0.9964 & 0.9942 & 0.9918 & 0.9856 & 0.9573 \\
& length & 5.8157 & 5.5294 & 5.3037 & 4.9166 & 4.0639 \\
\hline
\end{tabular}}
\caption{Coverage probabilities and average lengths of $\beta$-expectation tolerance intervals ($\beta=0.95$) for data generated from $N(0,1)$, based on 2,000 simulations for each sample size.}
\label{tab:sim_bayes_expect_tol_N}
\end{table}

\begin{table}[]
\centering
\resizebox{0.9\textwidth}{!}{
\begin{tabular}{l l | c c c c c} 
 Method &  & n=10 & n=30 & n=50 & n=100 & n=1000 \\
\hline
DP$(a=0)$
& coverage & 0.7969 & 0.9075 & 0.9292 & 0.9419 & 0.9491 \\
& length & 4.0448 & 5.3963 & 5.7746 & 5.9703 & 5.9881 \\
DP$(a=10, F_0=N(0,1))$
& coverage & 0.8614 & 0.9088 & 0.9259 & 0.9386 & 0.9487 \\
& length & 4.3511 & 5.2436 & 5.5872 & 5.8256 & 5.9716 \\
DP$(a=10, F_0=t_5)$
& coverage & 0.8981 & 0.9210 & 0.9324 & 0.9416 & 0.9490 \\
& length & 4.8965 & 5.5004 & 5.7576 & 5.9200 & 5.9824 \\
DP$(a=10, F_0=\text{Laplace}(0,1))$
& coverage & 0.9148 & 0.9277 & 0.9363 & 0.9435 & 0.9492 \\
& length & 5.2163 & 5.6603 & 5.8669 & 5.9831 & 5.9901 \\
DP$(a=100, F_0=N(0,1))$
& coverage & 0.8693 & 0.8867 & 0.8979 & 0.9138 & 0.9446 \\
& length & 4.0981 & 4.4188 & 4.6464 & 4.9945 & 5.8160\\
DP$(a=100, F_0=t_5)$
& coverage & 0.9268 & 0.9304 & 0.93303 & 0.9370 & 0.9474 \\
& length & 5.2606 & 5.3939 & 5.4876 & 5.6144 & 5.9189 \\
DP$(a=100, F_0=\text{Laplace}(0,1))$
& coverage & 0.9492 & 0.9488 & 0.9486 & 0.9482 & 0.9493 \\
& length & 5.9868 & 5.9986 & 6.0084 & 6.0021 & 5.9901 \\
\hline
\end{tabular}}
\caption{Coverage probabilities and average lengths of $\beta$-expectation tolerance intervals ($\beta=0.95$) for data generated from $\text{Laplace}(0,1)$, based on 2,000 simulations for each sample size.}
\label{tab:sim_bayes_expect_tol_L}
\end{table}

\clearpage
\subsubsection{Prior specification Using Empirical Bayes}
The above simulation studies demonstrate that choosing a suitable prior distribution can substantially improve the resulting tolerance limits when the sample size $n$ is small, even though mild prior misspecification is often not critical. While the MDP model generally performs well, it has high computational cost, and customizing its settings may require implementing a complex Gibbs sampler. A simple and practical alternative is to use the data to inform the prior distribution, following an empirical Bayes approach. 
To this end, we mimic the behavior of the posterior distributions of $a$ and $\xi$ in the MDP model. In particular, the posterior of $\xi|\bm{X}$ concentrates around the MLE of $\xi$ when $n$ is large. Regarding $a$, the properties of the Dirichlet Process (DP) imply that the posterior of $a|\bm{X}$ depends only on the partition structure of $\bm{X}$. When $\bm{X}$ corresponds to a continuous distribution, the probability of repeated values is essentially zero, so the partition structure depends only on $n$ and the hyperparameters $a_a$ and $b_a$. For fixed $a_a$ and $b_a$, the posterior mean of $a|\bm{X}$ is a concave increasing function of $n$ with very small curvature. In view of this, we take $\xi$ in the prior as its MLE, and $a$ as an increasing function of $n$, then apply the standard DP tolerance limit algorithm. Alternatively, one can elicit part of the parameters from prior knowledge and the remainder from the data.

Tables \ref{tab:sim_empbayes_expect_tol_N}–\ref{tab:sim_bayes_expect_tol2} illustrate the performance of this approach under various DGPs. For instance, in Table \ref{tab:sim_empbayes_expect_tol_N} (true $N(0,1)$ DGP), the first two rows correspond to $a \propto n$, the next two to $a \propto \sqrt{n}$, and the final two rows to using $a \propto \sqrt{n}$ with $\xi$ set to its MLE. Notably, setting $a \propto n$ performs poorly, which may seem at odds with the MDP results. In reality, this is consistent with the DP model; the posterior mean measure of a single DP combines the empirical distribution and the prior with weights $n/(a+n)$ and $a/(a+n)$, respectively. If $a \propto n$, the influence of the prior never diminishes, and any misspecification persists regardless of sample size, leading to a lack of robustness. Therefore, $a$ should increase at a rate strictly lower than $n$ to ensure that the prior effect vanishes as $n$ grows. This holds even if $\xi$ is chosen from the data; while a larger $n$ improves the MLE, a poorly specified prior can still degrade the tolerance limit if $a$ grows too quickly.

Guided by this principle, empirical Bayes strategies with $a \propto \sqrt{n}$ and $\xi$ set to its MLE provide reliable tolerance limits, except when $n$ is very small and the MLE is unstable. As shown in Tables \ref{tab:sim_empbayes_expect_tol_N} and \ref{tab:sim_empbayes_expect_tol_L}, this approach performs well even for $n \approx 30$ and approaches the performance of larger $n$ by $n \approx 50$. Hence, it is a practical option when eliciting a prior is difficult. 

\begin{table}[h]
\centering
\resizebox{0.9\textwidth}{!}{
\begin{tabular}{l l | c c c c c} 
Method &  & n=10 & n=30 & n=50 & n=100 & n=1000 \\
\hline
DP$(a=0.3n, F_0=t_5)$
& coverage & 0.8938 & 0.9440 & 0.9528 & 0.9581 & 0.9620 \\
& length & 3.4334 & 3.9606 & 4.0730 & 4.1328 & 4.1584 \\
DP($a=0.3n,F_0=\text{Laplace}(0,1))$
& coverage & 0.9106 & 0.9550 & 0.9621 & 0.9661 & 0.9690 \\
& length & 3.5958 & 4.1432 & 4.2539 & 4.3067 & 4.3228 \\
\hline
DP$(a=2\sqrt{n}, F_0=t_5)$
 & coverage & 0.9337 & 0.9490 & 0.9516 & 0.9535 & 0.9525 \\
& length & 3.8051 & 4.0292 & 4.0548 & 4.0532 & 3.9745 \\
DP$(a=2\sqrt{n},F_0=\text{Laplace}(0,1))$
 & coverage & 0.9522 & 0.9609 & 0.9606 & 0.9596 & 0.9544 \\
& length & 4.0802 & 4.2447 & 4.2250 & 4.1724 & 4.0104 \\
\hline
DP$(a=2\sqrt{n},F_0=\text{Laplace}(\mu, \sigma))$
& coverage & 0.8668 & 0.9305 & 0.9414 & 0.9483 & 0.9512\\
$\mu = \text{median}(\bm X),\, \sigma = \frac{1}{n}\sum|X_i-\mu|$ & length & 3.4157 & 3.8486 & 3.9330 & 3.9782 & 3.9522 \\
\hline
DP$(a=2\sqrt{n}, F_0 = N(\mu, \sigma))$
& coverage & 0.8322 & 0.9140 & 0.9303 & 0.9413 & 0.9490 \\
$\mu = \text{mean}(\bm{X}), \, \sigma=\text{sd}(\bm X)$, & length & 3.1236 & 3.6428 & 3.7736 & 3.8654 & 3.9153 \\
\hline
\end{tabular}}
\caption{Empirical Bayes $\beta$-expectation tolerance intervals under the true $N(0,1)$ DGP. $a$ is set as a function of $n$, and $\xi$ is set to its MLE. Results are based on 2,000 simulations for each setting with $\beta=0.95$.}
\label{tab:sim_empbayes_expect_tol_N}
\end{table}

\begin{table}[h]
\centering
\resizebox{0.9\textwidth}{!}{
\begin{tabular}{l l | c c c c c} 
 Method &  & n=10 & n=30 & n=50 & n=100 & n=1000 \\
\hline
DP$(a=0.3n,F_0=N(0,1))$
& coverage & 0.8356 & 0.9089 & 0.9241 & 0.9321 & 0.9363 \\
& length & 4.2232 & 5.2592 & 5.5009 & 5.5738 & 5.5282 \\
DP$(a=0.3n, F_0=t_5)$
& coverage & 0.8574 & 0.9201 & 0.9333 & 0.9406 & 0.9446 \\
& length & 4.4565 & 5.4946 & 5.7429 & 5.8287 & 5.8067 \\
\hline
DP$(a=2\sqrt{n},F_0=N(0,1))$
& coverage & 0.8514 & 0.9087 & 0.9245 & 0.9353 & 0.9463  \\
& length & 4.3046 & 5.2282 & 5.5177 & 5.6933 & 5.8777 \\
DP$a=2\sqrt{n},F_0=t_5)$
& coverage & 0.8818 & 0.9222 & 0.9332 & 0.9411 & 0.94803 \\
& length & 4.6987 & 5.6255 & 5.7460 & 5.8723 & 5.9439 \\
\hline
DP$(a=2\sqrt{n},F_0=\text{Laplace}(\mu, \sigma))$
& coverage & 0.8583 & 0.9222 & 0.9354 & 0.9436 & 0.9492 \\
$\mu = \text{median}(\bm X),\, \sigma = \frac{1}{n}\sum|X_i-\mu|$, & length & 4.7039 & 5.6255 & 5.8667 & 5.9800 & 5.9896 \\
\hline
DP$(a=2\sqrt{n}, F_0 = N(\mu, \sigma))$.
& coverage & 0.8367 & 0.9153 & 0.9312 & 0.9409 & 0.9483 \\
$\mu = \text{mean}(\bm{X}), \, \sigma=\text{sd}(\bm X)$, & length & 4.4538 & 5.4931 & 5.7620 & 5.8939 & 5.9533 \\
\hline
\end{tabular}}
\caption{Empirical Bayes $\beta$-expectation tolerance intervals under the true $\text{Laplace}(0,1)$ DGP. $a$ is set as a function of $n$ , and $\xi$ is set to its MLE. Results are based on 2,000 simulations for each setting with $\beta=0.95$.}
\label{tab:sim_empbayes_expect_tol_L}
\end{table}

\begin{table}[]
\centering
\resizebox{0.9\textwidth}{!}{
\begin{tabular}{l l | c c c c c} 
 &  & n=10 & n=30 & n=50 & n=100 & n=1000 \\
\hline
DP$(a=0)$
& coverage & 0.7858 & 0.8975 & 0.9207 & 0.9365 & 0.9485  \\
& length & 2.6147 & 3.3436 & 3.5344 & 3.6453 & 3.6624 \\
DP$(a=10,F_0=\text{Exp}(1))$
& coverage & 0.8985 & 0.9177 & 0.9282 & 0.9380 & 0.9481 \\
& length & 3.2525 & 3.4835 & 3.5842 & 3.6520 & 3.6623 \\
DP$(a=100,F_0=\text{Exp}(1))$
& coverage & 0.9443 & 0.9443 & 0.9446 & 0.9450 & 0.9483 \\
& length & 3.6580 & 3.6649 & 3.6667 & 3.6659 & 3.6624\\
DP$(a=0.3n,F_0=\text{Exp}(1))$
& coverage & 0.8490 & 0.9163 & 0.9310 & 0.9408 & 0.9485 \\
& length & 2.9162 & 3.4733 & 3.6012 & 3.6597 & 3.6624 \\
DP$(a=0.3n,F_0=\text{Gamma}(2,2))$
& coverage & 0.8131 & 0.8956 & 0.9139 & 0.9263 & 0.9362 \\
& length &  2.7276 & 3.2943 & 3.4212 & 3.4728 & 3.4654 \\
DP$(a=0.3n,F_0=\text{Half-Normal}(0,1))$
& coverage & 0.8207 & 0.9024 & 0.9198 & 0.9312 & 0.9400  \\
& length & 2.6624 & 3.2359 &  3.3633 & 3.4142 & 3.4073 \\
\hline
\end{tabular}}
\caption{Empirical Bayes $\beta$-expectation tolerance intervals under the true $\text{Exponential}(0,1)$ DGP. $a$ is set as a function of $n$, and $\xi$is set to its MLE. Results are based on 2,000 simulations for each setting with $\beta=0.95$.}
\label{tab:sim_bayes_expect_tol2}
\end{table}

\clearpage
\section{Relative Potency Data}
\label{sec:potency}
Having examined the performance of the proposed methods through simulation, we now turn to a real-world example. This case study analyzes relative potency data from drug manufacturing to illustrate how Bayesian tolerance intervals can support quality assurance decisions.

Ensuring consistent relative potency is essential in drug manufacturing, as it quantifies the biological effect of a test batch relative to a reference standard under identical conditions. Even small deviations from the target range can reduce efficacy or introduce safety risks.
Here, we analyze 25 relative potency measurements collected from five manufacturing campaigns of drug substance (Table~\ref{tab:data_potency}), as described in \cite{lewis:2022}.

\begin{table}[htb]
    \centering
    \begin{tabular}{c c c c c} 
    \hline
    Campaign 1 & Campaign 2 & Campaign 3 & Campaign 4 & Campaign 5 \\
    \hline
    95.661 & 98.830 & 96.665 & 98.198 & 95.922 \\
    102.259 & 94.887 & 106.234 & 98.186 & 102.956 \\
    103.135 & 103.362 & 103.735 & 107.872 & 101.596 \\
    99.827 & 94.117 & 104.317 & 99.987 & 96.806 \\
     & & 101.807 & 103.051 & 107.041 \\
     & & & 106.445 & 92.589 \\
    \hline
    \end{tabular}
    \label{tab:data_potency}
    \caption{Relative potency data (in units percent) from five manufacturing campaigns. Each campaign corresponds to a batch of drug substance.}
\end{table}
The values are scaled so that measurements between 90\% and 110\% of the reference potency meet quality specifications; values outside this range indicate potential manufacturing problems. Because relative potency varies from batch to batch, we can use tolerance intervals to determine whether nearly all future batches are likely to fall within these specifications. By comparing the limits of the intervals with the 90--110\% specification range, we can assess whether the manufacturing process is capable of consistently producing drug substances with acceptable potency. 

Table~\ref{tab:potency2Siid_tolint} reports the $(\beta=0.95, \gamma=0.95)$ two-sided tolerance intervals obtained under various prior distributions. For comparison, we also include Bayesian parametric tolerance intervals fitted using Howe’s method (HE) \citep{howe1969} for centered intervals and Owen’s approach (OCT) \citep{owen1964} for equal-tailed intervals under the $N(\mu, \sigma^2)$ likelihood model. Two prior settings are considered: 1) a non-informative prior with $p(\mu)\propto 1, p(\sigma)\propto 1/\sigma^2$, and 2) a conjugate prior $\mu|\sigma^2 \sim N(\mu_0, \sigma^2/n_0)$ and $1/\sigma^2 \sim \text{Gamma}(a_0, b_0)$, where $a_0=m_0/2$ and $b_0=m_0S_0^2/2$. We can interpret $n_0$ and $m_0$ as prior sample sizes for $\mu$ and $\sigma^2$, respectively, and $\mu_0$ and $S_0^2$ as the prior mean and variance, respectively. We set $\mu_0=100, n_0=5$, and $a_0=b_0=0.1$. Note that the non-informative setting corresponds to the frequentist method under a Gaussian assumption. All four fits yield tolerance intervals extending beyond the quality specification limits, suggesting that the process may not reliably produce batches within specifications.

For the Bayesian nonparametric models, we consider several DP and MDP formulations. For the DP model, we examine four prior settings with $a=1$ and $a=10$, using two base measures centered at 100 and spanning the [90, 110] specification limits by $2\sigma$ and $3\sigma$, i.e., $N(100,5^2)$ and $N(100,3.3^2)$, respectively. Under weaker priors and less dispersed base measures, the resulting tolerance intervals fall within the acceptable range, whereas tighter priors or stronger concentration ($a=10$) can push the limits outside the specifications. We also consider an MDP model with a vague prior $a \sim \text{Gamma}(1,1)$ and $F_0|\xi = N(\xi=(\mu_0, \sigma_0^2))$, with $\mu_0 \sim N(100,5^2)$ and $\sigma^2 \sim \text{InverseGamma}(1,1)$; this model produces limits that remain well within the acceptable range. Finally, several empirical Bayes configurations are explored, using Normal, Laplace, and Student-$t_5$ base measures centered at 100 with scales matched to the sample variance and concentration parameters $a=5=\sqrt{n}$ and $a=10=2\sqrt{n}$. These yield generally similar conclusions.

The results show that parametric Gaussian models produce conservative tolerance intervals extending beyond the target range, while the DP and MDP formulations adapt more flexibly to the empirical distribution of the data. This flexibility reduces the impact of the normality assumption and highlights that conclusions about process capability can depend strongly on model choice and prior dispersion.

\begin{table}
\centering
\resizebox{0.7\textwidth}{!}{
\begin{tabular}{l | c c} 
Method & Lower Limit & Upper Limit \\
\hline
HE Normal (noninformative prior) & 89.2396 & 111.9987 \\
HE Normal (conjugate prior) & 89.2317 & 111.8007 \\
\hline
OCT Normal (noninformative prior) & 88.40222 & 112.8361 \\
OCT Normal (conjugate prior) & 88.4591 & 112.5733 \\
\hline
DP$(a=1, F_0=N(100,3.3^2))$ & 92.5817 & 107.8836\\
DP$(a=10, F_0=N(100,3.3^2))$ & 92.1177 & 107.8836\\
DP$(a=1, F_0=N(100,5^2))$ & 92.1191 & 107.8762 \\
DP$(a=10, F_0=N(100,5^2))$ & 88.0546 & 111.9407 \\
\hline
MDP($a\sim\text{Gamma(1,1)},F_0 = N(\mu,\sigma^2))$ & 92.4686 & 109.4773 \\
 $\mu\sim N(100, 5^2),\,\sigma^2\sim \text{Inv-Gamma}(1,1)$& & \\
\hline
DP$(a=5, F_0=N(100,4.22^2))$ & 90.6887 & 109.2941 \\
DP$(a=5, F_0=\text{Laplace}(100,2.9847))$ &  89.2781 & 110.7232\\
DP$(a=5, F_0=3.2696t_5+100)$ & 89.9463 & 110.0551 \\
DP$(a=10, F_0=N(100,4.22^2))$ & 89.9092 & 110.0736 \\
DP$(a=10, F_0=\text{Laplace}(100,2.9847))$ & 87.8305 & 112.1709 \\
DP$(a=10, F_0=3.2696t_5+100)$ & 88.4986 & 111.5027 \\
\hline
\end{tabular}}
\caption{Two-sided $(\beta=0.95, \gamma = 0.95)$ tolerance intervals for the relative potency data, treating the 25 observations as i.i.d. Normal-based Bayesian intervals are shown first using the Howe (HE; centered) and Owen (OCT; equal-tailed) approaches with both noninformative and conjugate priors. The lower part of the table reports intervals obtained under DP and MDP priors for several choices of concentration parameter $a$ and base measure $F_0$.}
\label{tab:potency2Siid_tolint}
\end{table}


Table~\ref{tab:potency2Siid_exptolint} presents the corresponding $(\beta=0.95)$ expectation tolerance intervals under the same set of priors. All models conclude that the drug potency is within specifications. This reflects the less conservative nature of expectation tolerance intervals compared with $(\beta=0.95,\gamma)$ intervals with $\gamma \approx 1$, as they require only that a $\beta$ proportion of the population be covered on average rather than with near certainty. An additional observation is that the expectation tolerance intervals are remarkably similar across priors. This stability arises because the mean of the DP-based quantile process is less sensitive to tail behavior than its high-$\gamma$ quantiles. Consequently, expectation tolerance intervals provide a robust and practically interpretable summary of process variability.

Overall, this case study illustrates that Bayesian nonparametric tolerance intervals (particularly those based on the Dirichlet process) offer a flexible and data-adaptive framework for assessing manufacturing consistency. Compared with standard parametric approaches, they yield more realistic inferences about quality performance while remaining fully coherent within the Bayesian decision-making framework.

\begin{table}
\centering
\begin{tabular}{l | c c} 
Method & Lower Limit & Upper Limit \\
\hline
HE Normal (noninformative prior) & 91.8832 & 109.3551 \\
HE Normal (conjugate prior) &  91.8344 & 109.1979 \\
\hline
DP$(a=0)$ & 93.3948 & 107.4120 \\
DP$(a=1, F_0 = N(100,2^2))$ & 93.4317 & 107.4064 \\
DP$(a=10, F_0 = N(100,2^2))$ & 93.6685 & 107.2517 \\
DP$(a=1, F_0 = N(100,5^2))$ & 93.2753 & 107.4984\\
DP$(a=10,F_0 = N(100,5^2))$ & 92.4398 & 108.0114 \\
\hline
DP$(a=5, F_0 = N(100,4.22^2))$ & 93.1373 & 107.5455 \\
DP$(a = 5, F_0 = \text{Laplace}(100,2.9847)$ & 93.0554 & 107.6498 \\
DP$(a = 5, F_0 = 3.2696t_5+100)$ & 93.1061 & 107.6081 \\
DP$(a = 10, F_0 = N(100,4.22^2))$ & 92.9417 & 107.6361 \\
DP$(a = 10, F_0 = \text{Laplace}(100,2.9847)$ & 92.7944 & 107.8166 \\
DP$(a = 10, F_0 = 3.2696t_5+100)$ & 92.8869 & 107.7405 \\
\hline
\end{tabular}
\caption{Two-sided $\beta$-expectation tolerance intervals ($\beta = 0.95$) for the relative potency data, treating the 25 observations as i.i.d. Normal-based Bayesian intervals are shown first using the Howe (HE) approach with both noninformative and conjugate priors. The lower part of the table reports interval limits obtained under a variety of DP priors for different choices of concentration parameter $a$ and base measure $F_0$}
\label{tab:potency2Siid_exptolint}
\end{table}

\section{Discussion}
\label{sec:discussion}
We proposed a Bayesian method for constructing tolerance intervals using the Dirichlet process, offering a robust and flexible alternative to traditional parametric approaches. The method mitigates the consequences of misspecified distributional assumptions while preserving the ability to incorporate prior information, which is an advantage over frequentist nonparametric procedures, especially in small samples. Because the DP model remains analytically tractable, prior elicitation is straightforward and the resulting inference is computationally efficient, avoiding the need for MCMC.

Two decision-layer formulations provide flexibility in the conservativeness of the resulting intervals. Depending on the application, one may choose intervals based on guaranteeing a specified coverage probability or rely on the expectation-based decision rule, which typically performs well in practice. We recommend the expectation-based tolerance interval unless under-coverage carries unusually high cost.

We also explored a hierarchical extension through the MDP. While the MDP can improve performance in some settings, it comes at a cost in computational complexity. An empirical Bayes approximation offers a practical alternative when the simplicity of the pure DP approach is desired.

Several directions for extending Bayesian nonparametric tolerance intervals remain open. One challenge is the construction of two-sided $(\beta,\gamma)$ intervals under the DP. Deriving the required joint tail probabilities of two quantiles is conceptually straightforward but lacks a closed-form expression, making numerical evaluation and interval identification nontrivial. Relatedly, optimizing the length of two-sided intervals is considerably more difficult in the nonparametric setting, as the techniques available for parametric models do not transfer directly.

Finally, alternative nonparametric Bayesian priors, such as the Pitman–Yor process \citep{pitman1997two}, the Iswaran–James process \citep{ishwaran2001gibbs}, or dependent Dirichlet processes \citep{maceachern1999dependent}, could address some limitations of the DP, including its fixed relationship between variance and discreteness. However, these priors substantially increase the complexity of tracking the quantile process and would require more intensive computation. Incorporating such models into practical tolerance interval methodology is an interesting but challenging direction for future work.

\bibliographystyle{apalike}
\bibliography{references}

@article{aitchison1964two,
  title={Two papers on the comparison of {B}ayesian and frequentist approaches to statistical problems of prediction: Bayesian tolerance regions},
  author={Aitchison, J},
  journal={Journal of the Royal Statistical Society: Series B (Methodological)},
  volume={26},
  number={2},
  pages={161--175},
  year={1964},
  publisher={Wiley Online Library}
}

@article{antoniak1974mixtures,
  title={Mixtures of {D}irichlet processes with applications to {B}ayesian nonparametric problems},
  author={Antoniak, Charles E},
  journal={The annals of statistics},
  pages={1152--1174},
  year={1974},
  publisher={JSTOR}
}

@book{brent1973algorithms,
  author    = {Brent, Richard P.},
  title     = {Algorithms for Minimization without Derivatives},
  year      = {1973},
  publisher = {Prentice-Hall},
  address   = {Englewood Cliffs, NJ},
}

@book{casella2002statistical,
  author = {Casella, George and Berger, Roger L.},
  title = {Statistical Inference},
  year = {2002},
  edition = {2nd},
  publisher = {Duxbury},
  address = {Pacific Grove, CA},
}

@article{chen2022bayesian,
  title={Bayesian tolerance regions with an application to linear mixed models},
  author={Chen, X Gregory and van der Vaart, Aad},
  journal={arXiv preprint arXiv:2209.08496},
  year={2022}
}

@article{cho:2021,
  title={Tolerance intervals in statistical software and robustness under model misspecification},
  author={Cho, K. S. and Ng, H. K. T.},
  journal={Journal of Statistical Distributions and Applications},
  volume={8},
  number={10},
  pages={1--49},
  year={2021}
}

@article{dong:2015b,
  title={Using tolerance intervals for assessment of pharmaceutical quality},
  author={Dong, X. and Tsong, Y. and Shen, M. and Zhong, J.},
  journal={Journal of Biopharmaceutical Statistics},
  volume={25},
  number={2},
  pages={317--327},
  year={2015}
}

@article{dong:2015,
  title={Statistical considerations in setting product specifications},
  author={Dong, X. and Tsong, Y. and Shen, M.},
  journal={Journal of Biopharmaceutical Statistics},
  volume={25},
  number={2},
  pages={280--294},
  year={2015}
}

@techreport{fda2011process,
  author={FDA},
  title={Process Validation: General Principles and Practices},
  institution={U.S. Food and Drug Administration},
  year={2011}
}

@article{ferguson1973bayesian,
  title={A {B}ayesian analysis of some nonparametric problems},
  author={Ferguson, Thomas S},
  journal={The annals of statistics},
  pages={209--230},
  year={1973},
  publisher={JSTOR}
}

@article{ferguson1974prior,
  title={Prior distributions on spaces of probability measures},
  author={Ferguson, Thomas S},
  journal={The annals of statistics},
  pages={615--629},
  year={1974},
  publisher={JSTOR}
}

@book{ghosal2017fundamentals,
  title={Fundamentals of nonparametric {B}ayesian inference},
  author={Ghosal, Subhashis and van der Vaart, Aad W},
  volume={44},
  year={2017},
  publisher={Cambridge University Press}
}

@book{guttman1970statistical,
  title={Statistical Tolerance Regions: Classical and {B}ayesian},
  author={Guttman, Irwin},
  year={1970},
  publisher={Charles Griffin}
}

@article{hahn1970simple,
  title={A simple nonparametric tolerance limit procedure for large samples},
  author={Hahn, Gerald J.},
  journal={Technometrics},
  volume={12},
  number={3},
  pages={563--568},
  year={1970}
}

@book{hahn:meeker:2017,
  title={Statistical Intervals: A Guide for Practioners and Researchers, Second Edition},
  author={Meeker, W. Q. and Hahn, G. J. and Escobar, L. A.},
  year={2017},
  publisher={Wiley},
  address={New York}
}

@ARTICLE{hamada:2002,
  author =       {M. Hamada},
  title =        {Bayesian Tolerance Interval Control Limits for Attributes},
  journal =      {Quality and Reliability Engineering International},
  year =         {2002},
  volume =       {18},
  number = 		{1},
  pages =        {45--52},
}

@article{hamada2004bayesian,
  title={Bayesian prediction intervals and tolerance regions},
  author={Hamada, M. and Johnson, V. E. and Moore, L. M. and Wendelberger, J. R.},
  journal={Technometrics},
  volume={46},
  number={4},
  pages={452--461},
  year={2004}
}

@article{hanson2005bayesian,
  title={Bayesian nonparametric modeling and data analysis: an introduction},
  author={Hanson, Timothy E and Branscum, Adam J and Johnson, Wesley O},
  journal={Handbook of statistics},
  volume={25},
  pages={245--278},
  year={2005},
  publisher={Elsevier}
}

@incollection{hjort2007nonparametric,
  title={Nonparametric quantile inference using {D}irichlet processes},
  author={Hjort, Nils Lid and Petrone, Sonia},
  booktitle={Advances In Statistical Modeling And Inference: Essays in Honor of Kjell A Doksum},
  pages={463--492},
  year={2007},
  publisher={World Scientific}
}

@article{howe1969,
  title={Two-sided tolerance limits for normal populations --- Some improvements},
  author={Howe, W. G.},
  journal={Journal of the American Statistical Association},
  volume={64},
  pages={610--620},
  year={1969}
}

@techreport{ichq6a1999,
  author={ICH},
  title={{ICH Q6A}: Specifications: Test Procedures and Acceptance Criteria for New Drug Substances and New Drug Products: Chemical Substances},
  year={1999},
  institution={International Council for Harmonisation}
}

@article{ishwaran2001gibbs,
  title={Gibbs sampling methods for stick-breaking priors},
  author={Ishwaran, Hemant and James, Lancelot F.},
  journal={Journal of the American Statistical Association},
  volume={96},
  number={453},
  pages={161--173},
  year={2001},
  publisher={Taylor \& Francis}
}

@book{krishnamoorthy2009statistical,
  title={Statistical tolerance regions: theory, applications, and computation},
  author={Krishnamoorthy, Kalimuthu and Mathew, Thomas},
  year={2009},
  publisher={John Wiley \& Sons}
}

@article{krishnamoorthy2006improved,
  title={Improved tolerance intervals for normal populations},
  author={Krishnamoorthy, K. and Mondal, S.},
  journal={Technometrics},
  volume={48},
  number={1},
  pages={69--81},
  year={2006}
}

@incollection{lewis:2022,
  author      = "Lewis, R. and Hudson-Curtis, B.",
  title       = "Calculating statistical tolerance intervals using {SAS}",
  editor      = "Faya, P. and Pourmohamad, T.",
  booktitle   = "Case Studies in {B}ayesian Methods for Biopharmaceutical {CMC}",
  publisher   = "Taylor \& Francis",
  series      = "Chapman \& Hall/CRC Biostatistics Series",
  address     = "New York",
  year        = 2022,
  pages       = "266-290",
  chapter     = 14,
}

@article{little2016tolerance,
  author       = {Little, Thomas A.},
  title        = {Essentials in Tolerance Design and Setting Specification Limits},
  journal      = {BioPharm International},
  year         = {2016},
  volume       = {29},
  number       = {6},
  pages        = {41--45},
}

@inproceedings{maceachern1999dependent,
  title={Dependent nonparametric processes},
  author={MacEachern, Steven N.},
  booktitle={ASA Proceedings of the Section on Bayesian Statistical Science},
  pages={50--55},
  year={1999},
  organization={American Statistical Association}
}

@article{montes2019simple,
  title        = {Simple Approach to Calculate Random Effects Model Tolerance Intervals to Set Release and Shelf-Life Specification Limits of Pharmaceutical Products},
  author       = {Montes, Richard O. and Burdick, Richard K. and Leblond, David J.},
  journal      = {PDA Journal of Pharmaceutical Science and Technology},
  year         = {2019},
  volume       = {73},
  number       = {1},
  pages        = {39--59},
  doi          = {10.5731/pdajpst.2018.008839},
}

@article{muller2004nonparametric,
  author    = {M{\"u}ller, Peter and Quintana, Fernando A.},
  title     = {Nonparametric {B}ayesian data analysis},
  journal   = {Statistical Science},
  year      = {2004},
  volume    = {19},
  number    = {1},
  pages     = {95--110},
  doi       = {10.1214/088342304000000017}
}

@article{muller2013bayesian,
  author    = {M{\"u}ller, Peter and Mitra, Riten},
  title     = {Bayesian nonparametric inference--why and how},
  journal   = {Bayesian Analysis},
  year      = {2013},
  volume    = {8},
  number    = {2},
  pages     = {269--302},
  doi       = {10.1214/13-BA811}
}

@article{oliva:2025,
  title={Tolerance Intervals Under a Class of Unbalanced Linear Mixed Models},
  author={Oliva-Aviles, C. and Hauser, P.},
  journal={Technometrics},
  volume={67},
  pages={193--202},
  year={2025}
}

@article{owen1964,
  title={Controls of Percentages in Both Tails of the Normal Distribution},
  author={Owen, D. B.},
  journal={Technometrics},
  volume={6},
  pages={377--387},
  year={1964}
}

@article{pitman1997two,
  title={The two-parameter Poisson--Dirichlet distribution derived from a stable subordinator},
  author={Pitman, Jim and Yor, Marc},
  journal={Annals of Probability},
  volume={25},
  number={2},
  pages={855--900},
  year={1997},
  publisher={Institute of Mathematical Statistics}
}

@book{press2007numerical,
  author    = {Press, William H. and Teukolsky, Saul A. and Vetterling, William T. and Flannery, Brian P.},
  title     = {Numerical Recipes: The Art of Scientific Computing},
  edition   = {3rd},
  year      = {2007},
  publisher = {Cambridge University Press},
  address   = {Cambridge, UK},
}

@article{schwenke:2021,
  title={A practical discussion on estimating shelf life through tolerance intervals},
  author={Schwenke, J. and Forenzo, P. and Stroup, W. and Quinlan, M.},
  journal={AAPS PharmSciTech},
  volume={273},
  number={22},
  pages={1--12},
  year={2021}
}

@article{tendong2020ac50,
  title        = {Controlling the Reproducibility of {AC50} Estimation during Compound Profiling through {B}ayesian $\beta$-Expectation Tolerance Intervals},
  author       = {Tendong, W. and Lebrun, P. and Verbist, B.},
  journal      = {SLAS Discovery},
  year         = {2020},
  volume       = {25},
  number       = {9},
  pages        = {1009--1017},
  doi          = {10.1177/2472555220918201},
}

@article{walker1999bayesian,
  title={Bayesian nonparametric inference for random distributions and related functions},
  author={Walker, Stephen G and Damien, Paul and Laud, Purushottam W and Smith, Adrian FM},
  journal={Journal of the Royal Statistical Society: Series B (Statistical Methodology)},
  volume={61},
  number={3},
  pages={485--527},
  year={1999},
  publisher={Wiley Online Library}
}

@article{wilks1941determination,
  title={Determination of sample sizes for setting tolerance limits},
  author={Wilks, Samuel S.},
  journal={Annals of Mathematical Statistics},
  volume={12},
  number={1},
  pages={91--96},
  year={1941}
}

@article{young2014nonparametric,
  title={Nonparametric tolerance limits and confidence limits for percentiles},
  author={Young, D. S. and Mathew, T.},
  journal={Journal of Statistical Computation and Simulation},
  volume={84},
  number={5},
  pages={1003--1016},
  year={2014}
}

\end{document}